\documentclass[pdflatex,sn-mathphys-num]{sn-jnl}

\usepackage{graphicx}%
\usepackage{multirow}%
\usepackage{amsmath,amssymb,amsfonts}%
\usepackage{amsthm}%
\usepackage{mathrsfs}%
\usepackage[title]{appendix}%
\usepackage{xcolor}%
\usepackage{textcomp}%
\usepackage{manyfoot}%
\usepackage{booktabs}%
\usepackage{algorithm}%
\usepackage{algorithmicx}%
\usepackage{algpseudocode}%
\usepackage{listings}%

\theoremstyle{thmstyleone}%
\theoremstyle{thmstyletwo}%

\theoremstyle{thmstylethree}%

\begin{document}

\title[Cluster-based Structural Similarity for Dataset Visualization and Data Selection for Machine Learning Interatomic Potentials]{Cluster-based Structural Similarity for Dataset Visualization and Data Selection for Machine Learning Interatomic Potentials}


\author*[1,2]{\fnm{Yuto} \sur{Iwasaki}}\email{iwasaki.yuto@fujitsu.com}

\author[2]{\fnm{Miguel A.} \sur{Caro}}\email{miguel.caro@aalto.fi}

\affil*[1]{\orgdiv{Fujitsu Research}, \orgname{Fujitsu Limited}, \orgaddress{\street{4-1-1 Kamikodanaka}, \city{Kawasaki}, \postcode{211-8588}, \state{Kanagawa}, \country{Japan}}}

\affil[2]{\orgdiv{School of Chemical Engineering}, \orgname{Aalto University}, \orgaddress{\street{Otakaari 1 B}, \city{Espoo}, \postcode{02150}, \country{Finland}}}


\abstract{Machine learning interatomic potentials (MLIPs) are essential components for accelerating simulation-driven materials design. Data-efficient MLIP training relies on data-selection strategies that maximize structural diversity while limiting computationally expensive first-principles calculations. A key challenge in such strategies is evaluating structural similarity, which involves a trade-off between retaining information on individual atomic environments and reducing computational cost. Here, we propose similarity evaluation methods that achieve both representational fidelity and computational efficiency. Our method represents each structure using a small set of characteristic atomic environments identified by $k$-medoids clustering and computes pairwise similarity through optimal matching between these representatives or their distributions. Molecular benchmarks demonstrate that our method is approximately $50$ times faster than the baseline method while also more clearly distinguishing structures with different chemical compositions. Similarity-based data-selection benchmarks demonstrate that our methods improve the data efficiency and stability of force prediction in MLIPs.}




\maketitle

\section{Introduction}\label{introduction}
Recent advances in computational power and simulation techniques enable the accurate prediction of material properties, facilitating the simulation-driven design of novel materials. Molecular dynamics (MD) simulation, a method that computes the temporal evolution of atoms according to the equations of motion, is widely used for investigating material behavior and performance, as well as chemical reaction mechanisms.

The accuracy of MD simulations is fundamentally dependent on the method used to compute the interatomic forces. While first-principles calculations, typically based on density functional theory (DFT), can achieve high accuracy, their significant computational cost severely limits the number of atoms and the time scales to be simulated. To overcome this limitation, researchers have turned to machine learning interatomic potentials (MLIPs) to predict these forces \cite{behler2007generalized,bartok2010gaussian}. MLIPs are surrogate models, typically based on neural networks or Gaussian process regression, trained to reproduce potential energies and atomic forces obtained from DFT calculations. Well-trained MLIPs achieve prediction accuracy comparable to DFT while being orders of magnitude faster. The broad applicability of MLIPs has been demonstrated through MD simulations of diverse systems, including inorganic materials \cite{deringer2017machine,ibragimova2025unifying,kang2025anomalous,ito2025predicting} and molecular systems \cite{yoshimoto2026active,mohanty2023development}. This computational efficiency makes it feasible to perform large-scale and long-duration MD simulations, thereby facilitating the precise characterization of diverse properties for a wider range of materials.

The performance of MLIPs is dictated by both their architecture and the quality of their training data. Various architectures have been proposed, each modeling atomic interactions through different approaches. The Behler-Parrinello neural network pioneered this approach by combining neural networks with hand-crafted and atom-centered descriptors that are invariant under translation, rotation, and permutation of chemically equivalent atoms \cite{behler2007generalized}. Another prominent approach is the Gaussian Approximation Potential (GAP), which combines an invariant representation of the local atomic environment with Gaussian process regression \cite{bartok2010gaussian}. Moment Tensor Potentials were introduced, combining moment tensor descriptors and linear regression to improve computational efficiency \cite{shapeev2016moment}. The next generation of MLIPs moved away from hand-crafted descriptors, instead learning representations of atomic environments directly from data using neural networks. A prime example is the Deep Potential model, which employs two interconnected neural networks: one to construct a learnable descriptor from the atomic environment, and another to map this descriptor to the energies and forces, thereby achieving high expressive power \cite{wang2018deepmd,zeng2023deepmd,zeng2025deepmd}. Graph neural network-based models have become increasingly prominent. Models such as NequIP \cite{batzner20223}, Allegro \cite{musaelian2023learning}, PaiNN \cite{schutt2021equivariant}, EquiformerV2 \cite{liao2024equiformerv2}, and MACE \cite{batatia2022mace} achieve high performance on benchmark datasets. More recently, the research field has shifted from domain-specific MLIPs toward foundation MLIPs pretrained on large and diverse datasets, including CHGNet \cite{deng2023chgnet}, MACE-MP-0 \cite{batatia2025foundation}, MatterSim \cite{yang2024mattersim}, and UMA \cite{wood2026family}. Meanwhile, self- and semi-supervised learning techniques are also being explored to enable more data-efficient pretraining of foundation MLIPs \cite{oyama2025lamm,majima2025self}. Recent studies have explored knowledge distillation from accurate but computationally expensive foundation MLIPs to faster domain-specific MLIPs, demonstrating its potential for the data-efficient construction of MLIPs \cite{zhang2024dpa,matsumura2025knowledge}. Despite these advances in model architecture and training strategies, MLIP accuracy and robustness remain critically dependent on the coverage and quality of the training dataset.

Constructing a high-quality training dataset is therefore another central challenge in MLIP development. Active learning has emerged as a powerful approach for addressing this challenge. This method consists of an iterative cycle of training the MLIP and adding new data to the training set. Since the DFT calculations required to label data are computationally expensive, it is crucial to strategically select new data. This process must avoid adding redundant data that is similar to the existing training dataset and instead select data that contributes most to improving the MLIP's prediction accuracy. Various selection strategies have been proposed \cite{zhang2020dp,vandermause2020fly,van2023hyperactive,matsumura2025generator}, primarily based on uncertainty of the MLIP's predictions or structural diversity, while some approaches incorporate domain-specific criteria into data selection \cite{yamazaki2025improving}.

Uncertainty-based approaches utilize metrics such as the variance in predictions from an ensemble of MLIPs \cite{smith2018less,zhang2020dp,matsumura2025generator}, the predictive variance from Gaussian process regression \cite{vandermause2020fly}, and various other techniques \cite{tan2023single}. Crucially, these uncertainty-based approaches are not mutually exclusive with the structural-diversity-based ones. Therefore, to isolate and quantify the specific contribution of similarity-based approaches, this study focuses exclusively on them.

Structural-diversity-based approaches represent each structure as a feature vector and sample diverse structures in the resulting feature space, sometimes after dimensionality reduction. While this approach is effective for obtaining a global overview of the dataset's diversity, it faces challenges in the step of converting data into a single vector representation. For example, the strategy of concatenating all atomic feature vectors into a single structure-level vector is constrained by the requirement that the number of atoms must be constant across the dataset \cite{shimizu2021phase}. Alternatively, averaging all atomic feature vectors of each structure into a single representative vector sacrifices representation fidelity, as information about diverse or unique local environments may be lost in this process \cite{matsumura2025generator}. A more sophisticated approach, which utilizes the output from a pre-trained Graph Neural Network like M3GNet \cite{chen2022universal}, generates a more informative representation, but it also condenses all structural information into a single feature vector, and therefore, information about the diversity of internal atomic environments is compressed and potentially lost \cite{qi2024robust}.

Our work is motivated by the limitations of data-selection strategies that rely on comparing compressed, single-vector representations. We argue for a more direct approach: data-selection strategies based on pairwise comparisons between atomic structures. By directly assessing the similarity between any data pair, we can more effectively facilitate the selection of structurally diverse configurations, thereby enhancing the overall diversity of the training dataset. The success of this framework, however, is critically dependent on the method used to compute the similarity between data pairs.

Comparing the feature vectors of two individual atoms is straightforward, as their similarity can be directly evaluated using measures such as the inner product. Similarly, comparing two structures each represented by a single global feature vector is also computationally simple. In contrast, directly comparing two structures containing different numbers of atoms is substantially more challenging, as each structure is represented by a variable-sized set of feature vectors rather than a single vector. Consequently, similarity evaluation requires determining how the feature vectors of the two structures should be compared and matched.

An existing approach for computing a similarity score is to directly compare the entire set of atomic feature vectors from two structures \cite{de2016comparing}. While this method can yield a high-fidelity similarity score, the associated computational cost of solving a combinatorial optimization problem for each pair is extremely high, which limits its scalability and practical application to large datasets.

The objective of this study is to develop a structure-level similarity evaluation method that balances representational fidelity and computational cost. Specifically, our method represents each structure with a small set of representative vectors using $k$-medoids clustering and calculates pairwise similarity based on these representative vectors. In this paper, we first qualitatively evaluate the ability to distinguish between chemically distinct molecules. Subsequently, we integrate the similarity evaluation method into a farthest-point-sampling (FPS)-based data selection framework to verify its practical effectiveness in improving the data selection efficiency for training MLIPs.


\section{Results}
We evaluate the proposed similarity evaluation methods from two perspectives. The first concerns their ability to distinguish chemically distinct data points, thereby assessing their fundamental discriminative capability. The second concerns their practical utility in data selection, for which we examine whether a data selection framework employing the similarity evaluation method can contribute to the data-efficient training of MLIPs.

\subsection{Overview of Similarity Evaluation Methods}
Typically, MLIPs can be understood as a two-stage pipeline. In the first stage, a descriptor transforms the local atomic environments, including positions and chemical species of each atom relative to surrounding atoms, into a fixed-size numerical feature vector. In the second stage, a fitting model maps these feature vectors to physical properties such as potential energies and atomic forces. The similarity evaluation methods discussed in this section operate on the set of atomic feature vectors generated in the first stage, which encode the chemical and structural information used to predict physical properties.

A primary challenge arises when comparing two data points with differing numbers of atoms. Each data point is represented by a set of feature vectors, with one vector for each of its atoms. Consequently, a difference in atom count leads to sets of varying sizes between both data points, making a direct comparison non-trivial. The final goal of the comparison is to reduce this complex relationship into a scalar score representing their similarity. Four distinct strategies emerge to address this challenge:

\begin{enumerate}
\item Data-wise similarity evaluation method: An aggregation-based strategy summarizes the set of feature vectors for each data point into a single representative vector before comparison. A similarity score is then computed based on these two vectors. While this approach is computationally efficient, its fidelity is constrained because the diversity of atomic environments is lost in the single-vector representation.
\item Atom-wise similarity evaluation method: A direct strategy computes a similarity score by performing a comprehensive comparison between all feature vectors of the two data points. This strategy determines an optimal correspondence between the two sets of feature vectors, even when the data points contain different numbers of atoms. In contrast to the data-wise method, this method provides an exhaustive and high-fidelity comparison, but at a significant computational expense due to the computational complexity of this matching problem.
\item Cluster-wise similarity evaluation method: Our proposed strategy first selects multiple representative vectors for each data point. A similarity score is then computed by performing a comparison between sets of representative vectors. This approach is designed to be a balanced, intermediate solution that bridges the gap between the computational efficiency of the data-wise method and the representational fidelity of the atom-wise method.
\item Distribution-aware cluster-wise similarity evaluation method: This extension initially follows the same procedure as the cluster-wise method to select multiple representative vectors for each data point. It then incorporates both the representative vectors and the distribution of feature vectors within each cluster into the calculation of a data-level distance. Although this additional information increases the computational cost relative to the standard cluster-wise method, the cost is expected to remain lower than that of the atom-wise method.
\end{enumerate}

\subsection{Qualitative Validation of Similarity}
\subsubsection{Analysis Procedure}
We first perform a qualitative validation to assess our method's core function: its ability to capture chemically meaningful relationships. A robust and accurate method should assign high similarity scores to similar data pairs and low scores to dissimilar ones. Accordingly, we test this capability by applying our cluster-wise methods and the baseline atom-wise and data-wise methods to a dataset comprising various chemical species, compositions, and structures.

For the three similarity-based methods, the evaluation pipeline consists of four stages.
\begin{enumerate}
\item The process begins by computing a similarity matrix $\mathbf{K}$ for all data pairs using three similarity evaluation methods.
\item Each similarity score in this matrix $\mathbf{K}$ is then transformed into a Euclidean distance using the equation below, yielding a distance matrix $\mathbf{D}$,
\begin{equation}
d_{\mathrm{A},\mathrm{B}}=\sqrt{2-2{k_{\mathrm{A,B}}}^{1/\zeta}}.
\end{equation}
Note that since the similarity scores lie in the range from $0$ to $1$, the distance scores are also bounded by 0 and $\sqrt{2}$.
\item   Subsequently, we apply multidimensional scaling (MDS) or isometric feature mapping (Isomap) \cite{tenenbaum2000global}, which are dimensionality reduction algorithms, to $\mathbf{D}$ to make a scatter plot visualizing the dissimilarity within the dataset.
\item  In the final stage, the data points in the scatter plot are colored by their chemical compositions, and the resulting separation is visually inspected.
\end{enumerate}

Note that, for the distribution-aware cluster-wise method, the Wasserstein distance is intrinsically a distance measure. Therefore, it is used directly without constructing a similarity matrix or applying the similarity-to-distance conversion.

This visualization enables a qualitative evaluation of the similarity evaluation methods. Although other properties, such as the potential energy, could also be used for visualization, we chose chemical composition because it is directly related to the design of the similarity metrics. In both the atom-wise and cluster-wise methods, the similarity between feature vectors belonging to different chemical species is defined to be zero. Consequently, data points with different compositions are expected to be well separated in the resulting embedding. By coloring the data points in the scatter plot according to their composition, we can therefore visually verify whether the spatial arrangement of the points reflects their chemical similarities. 

Although alternative visualization methods for the distance matrix, such as UMAP \cite{mcinnes2018umap}, t-SNE \cite{van2008visualizing}, and clMDS \cite{hernandez2024cluster}, could also be employed, we adopted MDS and Isomap because they directly embed the distance matrix produced by the similarity evaluation methods with simple mathematics. This property makes the resulting visualization particularly suitable for assessing whether the computed distances reflect meaningful chemical similarities.

\subsubsection{Target Datasets}
For this validation, we use two datasets. One is the QM7b dataset \cite{blum2009970,montavon2013machine}, which consists of small organic molecules. Since this dataset contains a diverse set of chemical species (H, C, N, O, S, and Cl), it serves as an ideal benchmark for evaluating whether a method can capture compositional differences. Because the number of representative vectors is set to five in this validation, we exclude one molecule containing only four atoms. The final dataset used in this analysis therefore comprises $7,210$ molecules, with the number of atoms per data point ranging from $5$ to $23$.

\begin{figure}
\includegraphics[width=\textwidth]{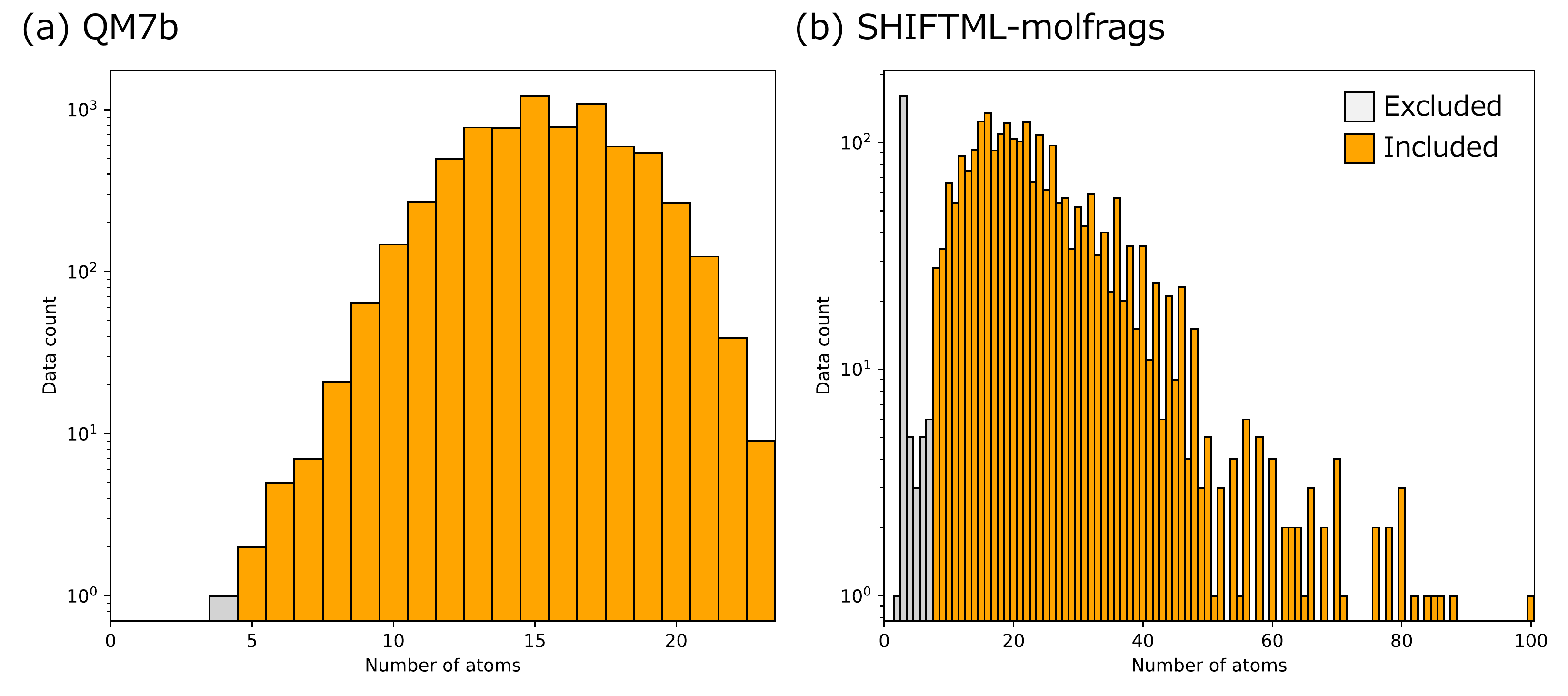}
\caption{Distribution of the number of atoms in the (a) QM7b and (b) SHIFTML-molfrags datasets before and after filtering.} 
\label{fig:dataset}
\end{figure}

The other is the SHIFTML-molfrags subset of the PET-MAD dataset \cite{mazitov2025pet}, which consists of neutral molecular fragments. This dataset contains a diverse set of chemical species (H, C, N, O, and S). Because molecules with very few atoms are relatively easy to compare, data points containing fewer than eight atoms were excluded, with the threshold chosen to remove simple molecules while avoiding a substantial reduction in dataset size. The final dataset used in this analysis comprises $2,592$ molecules, with the number of atoms per data point ranging from $8$ to $100$. Owing to its substantially broader distribution of the number of atoms than that of the QM7b dataset as shown in Fig.~\ref{fig:dataset}, this dataset presents a more challenging benchmark for similarity evaluation.

\subsubsection{Validation on QM7b Dataset}
\begin{figure}
\includegraphics[width=\textwidth]{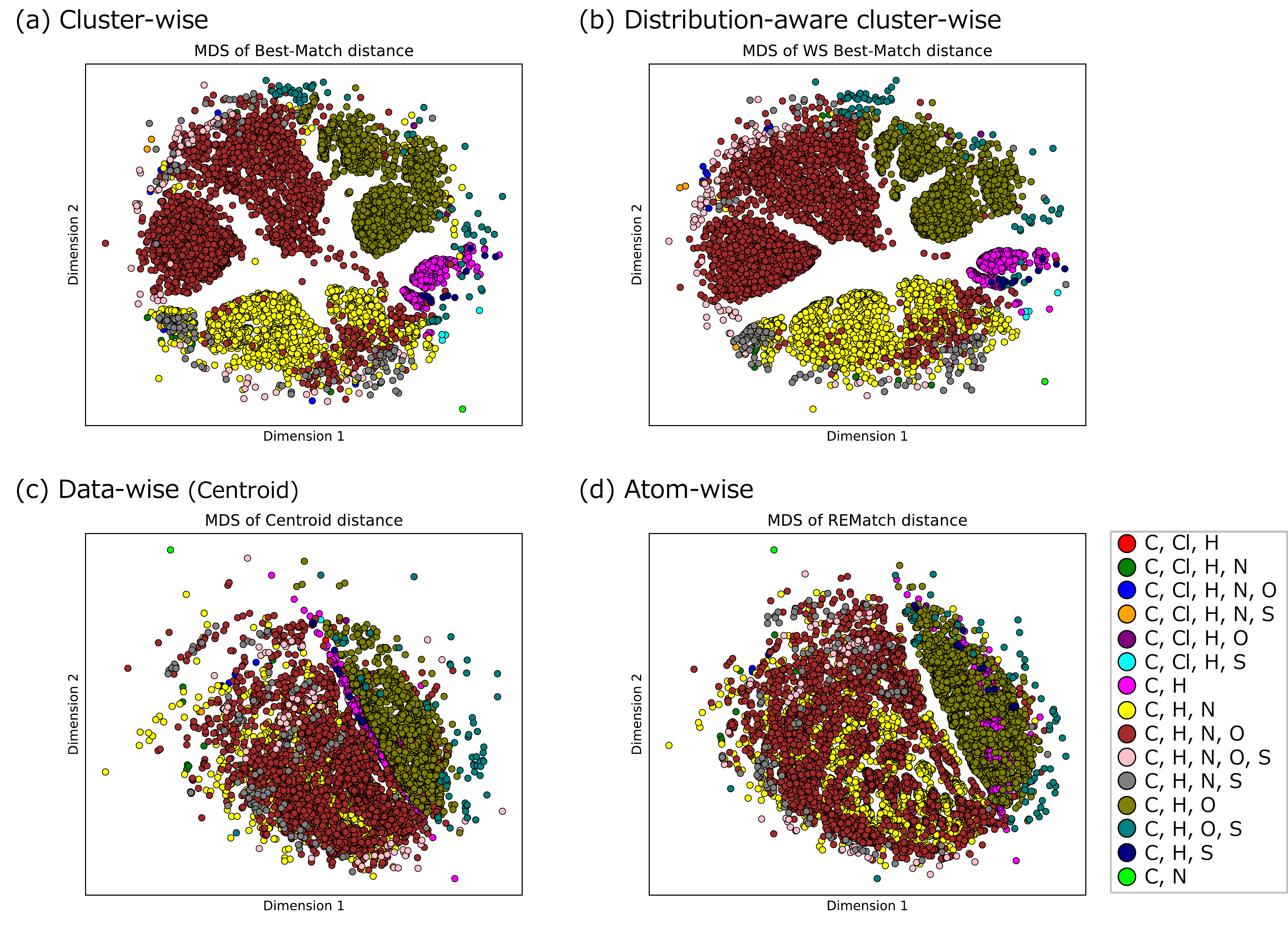}
\caption{Comparison of the MDS visualizations of the distance matrices derived from the (a) cluster-wise, (b) distribution-aware cluster-wise, (c) data-wise, and (d) atom-wise methods. The analysis is performed on the QM7b dataset. Each point represents a data point and is colored according to its chemical composition.}
\label{fig:result_qm7b}
\end{figure}

Fig.~\ref{fig:result_qm7b} shows MDS visualizations of the distance matrices derived from the proposed and baseline methods for the QM7b dataset. Each point represents a data point and is colored according to its chemical composition. The cluster-wise method includes a key hyperparameter: the number of representative vectors, which is set to $5$ in this validation. Detailed conditions are provided in Appendix~\ref{appendix:qm7b}.

The results show that both cluster-wise methods form clearly separated groups corresponding to the compositions in the dataset. The atom-wise method shows less distinct separation, resulting in a partial intermingling of data with different compositions. The data-wise method produces a single, undifferentiated cluster. These results suggest that both cluster-wise methods have the ability to reflect compositional differences, including those arising from the presence or absence of a single chemical species, in their similarity scores.

This performance gap stems from a fundamental difference in the level of information aggregation. The atom-wise method compares all atoms between data pairs. In the QM7b dataset, light hydrogen atoms are predominant compared to the other chemical species. Consequently, the similarity calculation shown in Eq.~\eqref{atom_wise:k} is dominated by the comparisons between these majority hydrogen atoms. As a result, a chemically meaningful difference, such as the presence or absence of a single heavy atom, is diluted by the numerous hydrogen atom comparisons, and its contribution to the overall similarity score is diminished. This is considered the reason for the insufficient separation by composition. On the other hand, the data-wise method loses too much information by compressing everything into a single vector, resulting in insufficient separation.

In contrast, the proposed cluster-wise method first summarizes the atomic environments into a few representative vectors identified through clustering. In this process, even an atom that occurs only once in a data point can form its own cluster and be selected as a representative vector, provided that its local environment is distinct. During the comparison, these few representative vectors are compared against each other, meaning the similarity scores between distinct representative vectors have a significant influence on the overall similarity score as shown in Eq.~\eqref{cluster_wise:k}. In other words, the clustering process increases the relative contribution of atomic species that are chemically important but numerically in the minority.


Beyond its discriminative capability, the cluster-wise method offers a significant advantage in terms of computational efficiency. Table~\ref{tab:time} compares the processing times of the cluster-wise, distribution-aware cluster-wise, and atom-wise methods. The results confirm that our proposed method achieves approximately a $50$-fold speedup over the atom-wise method. The primary factor behind this is the number of representative vectors involved in the similarity calculation. In the atom-wise method, all atoms in a data point serve as representative vectors, so the number of vectors to be compared depends on the system size, ranging from $5$ to $23$ in this dataset. In contrast, our cluster-wise method summarizes the atomic environments via clustering, reducing the comparison to a fixed set of $5$ representative vectors. This dramatically reduces the scale of the optimal matching problem.

Although the distribution-aware cluster-wise method is approximately one order of magnitude slower than the standard cluster-wise method because of the additional Wasserstein distance calculations, it remains substantially faster than the atom-wise method. This result confirms that the proposed extension preserves a significant computational advantage while accounting for the distribution of feature vectors within each cluster.
\begin{table}
\caption{Comparison of processing time for cluster-wise, distribution-aware cluster-wise, and atom-wise methods. The table lists the mean processing time and corresponding standard deviation, calculated from 5 independent runs.}
\label{tab:time}
\begin{tabular}{cc}
\hline
Method & Mean and standard deviation of processing time\\
\hline
Cluster-wise method & $2.43\times10^3 \pm 4.10\times10$ s\\
Distribution-aware cluster-wise method&$1.95\times10^4 \pm 1.93\times10^2$ s\\
Atom-wise method & $1.19\times10^5 \pm 2.7\times 10^4$ s\\
\hline
\end{tabular}
\end{table}

\subsubsection{Validation on SHIFTML-molfrags Dataset}
\begin{figure}
\includegraphics[width=\textwidth]{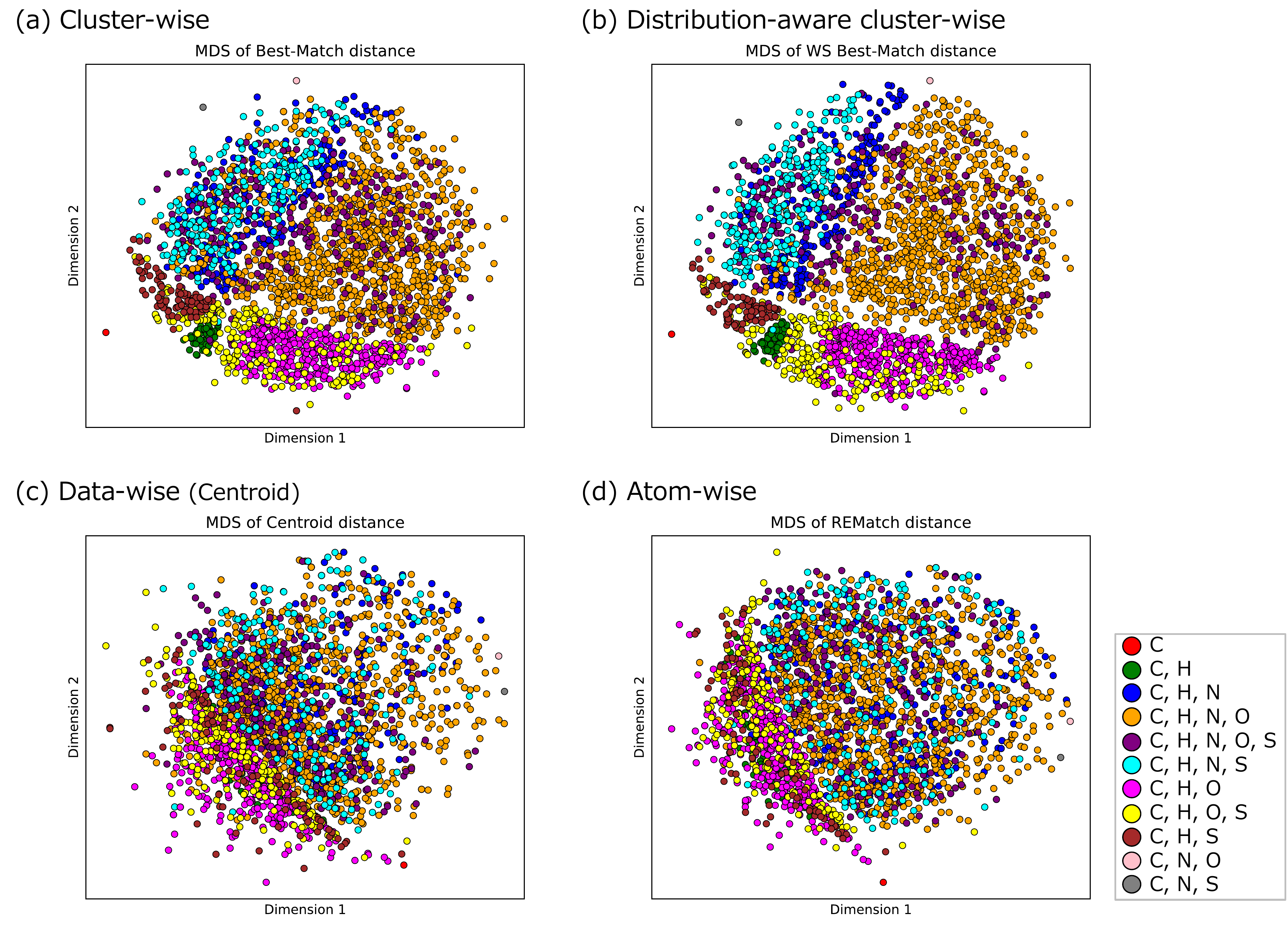}
\caption{Comparison of the MDS visualizations of the distance matrices derived from (a) the cluster-wise, (b) distribution-aware cluster-wise, (c) data-wise, and (d) atom-wise methods. The analysis is performed on the SHIFTML-molfrags dataset. Each point represents a data point and is colored according to its chemical composition.}
\label{fig:result_molfrags_mds}
\end{figure}

Fig.~\ref{fig:result_molfrags_mds} shows MDS visualizations of the distance matrices derived from the proposed and baseline methods for the SHIFTML-molfrags dataset. The number of representative vectors of the cluster-wise method is set to $6$ in this validation. Detailed conditions are shown in Appendix~\ref{appendix:shiftml}.

Similar to the results obtained for the QM7b dataset, both cluster-wise methods produced a clear separation of molecular compositions, whereas the data-wise and atom-wise methods showed substantially poorer separation. 

A particularly notable feature of the cluster-wise methods is the clear separation between compositions containing nitrogen and those without nitrogen. Furthermore, within each of these two groups, a gradual ordering according to chemical composition can be observed. This indicates that the proposed methods capture not only coarse compositional differences but also finer variations in chemical composition.

Unlike the QM7b dataset, the SHIFTML-molfrags dataset exhibits a much broader range of molecular sizes and greater structural diversity as shown in Fig.~\ref{fig:dataset}. Consequently, the resulting distance relationships are expected to be more complex. As MDS directly embeds the distance matrix into a low-dimensional Euclidean space, it may not fully preserve such complex relationships. Therefore, to further evaluate the robustness of the observed trends, we additionally employed Isomap, which preserves geodesic distances on a neighborhood graph and can better represent nonlinear distance relationships.

\begin{figure}
\includegraphics[width=\textwidth]{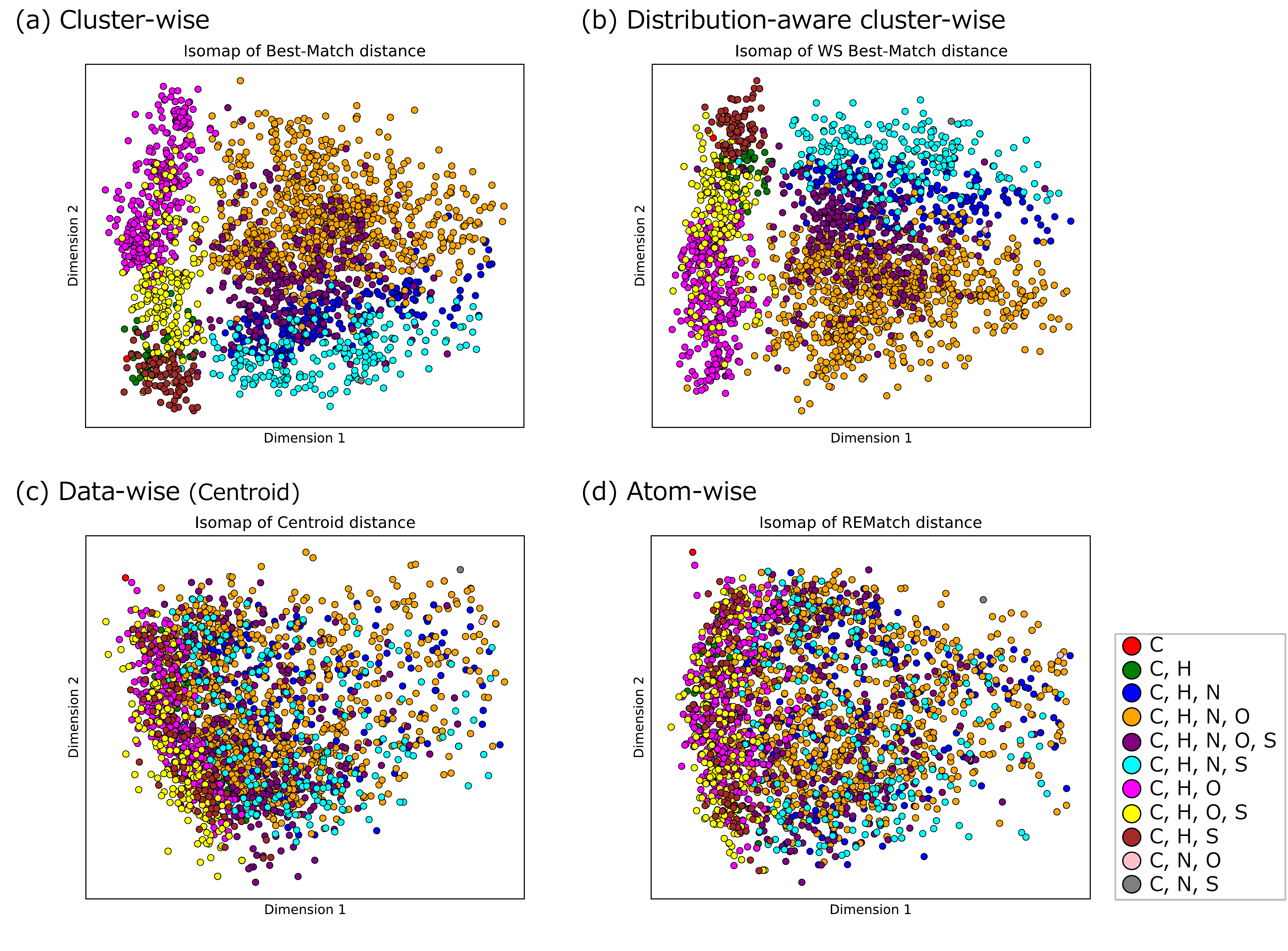}
\caption{Comparison of the Isomap visualizations of the distance matrices derived from (a) the cluster-wise, (b) distribution-aware cluster-wise, (c) data-wise, and (d) atom-wise methods. The analysis is performed on the SHIFTML-molfrags dataset. Each point represents a data point and is colored according to its chemical composition.}
\label{fig:result_molfrags_isomap}
\end{figure}

Fig.~\ref{fig:result_molfrags_isomap} shows Isomap visualizations of the distance matrices derived from the proposed and baseline methods for the SHIFTML-molfrags dataset, using $15$ nearest neighbors. The Isomap visualizations exhibit trends that are highly consistent with those observed using MDS. Both cluster-wise methods again produce substantially clearer separation of chemical compositions than the data-wise and atom-wise methods. In particular, the separation between compositions with and without nitrogen remains evident, together with the gradual compositional ordering observed within each group.

The consistency between the MDS and Isomap results is notable. Because these visualization methods rely on fundamentally different embedding principles, obtaining similar compositional separation in both cases suggests that the distance matrices produced by the proposed methods intrinsically capture chemically meaningful relationships among molecules, rather than reflecting an effect of the visualization method. This observation indicates that the clear separation achieved by the proposed methods originates from the quality of the underlying distance matrices rather than from the choice of visualization method.

\subsubsection{Summary}
The qualitative validation results demonstrate that the proposed cluster-wise methods consistently outperform both the atom-wise and data-wise methods in distinguishing chemically distinct molecules. This trend is observed for both the QM7b dataset, which contains molecules with $5$ to $23$ atoms, and the SHIFTML-molfrags dataset with a substantially broader size range of $8$ to $100$ atoms. These results suggest that the cluster-wise methods provide an effective framework for visualizing the relationships between molecules with varying numbers of atoms.

In addition, the cluster-wise method provides a computational advantage. For the QM7b dataset, the cluster-wise method achieved an approximately 50-fold reduction in computational time compared with the atom-wise method.

These results suggest that the cluster-wise framework effectively bridges the gap between atom-wise and data-wise representations. By grouping atoms with similar local environments through clustering and representing each molecule using a small number of representative vectors, the proposed methods preserve chemically meaningful information while suppressing redundant contributions. This enables the proposed methods to avoid both the excessive compression of the data-wise approach and the high computational cost of the atom-wise approach.

Another practical advantage of the cluster-wise method is its simplicity. In the atom-wise method, matching molecules with different numbers of atoms requires using the REMatch method, which introduces several hyperparameters, such as the regularization coefficient, the maximum number of iterations, and the convergence criterion. In contrast, the cluster-wise method represents each molecule using the same predefined number of clusters, making the matching problem fixed in size. As a result, the framework is primarily controlled by a single hyperparameter: the number of clusters. The proposed framework also allows meaningful visualizations to be obtained without more specialized visualization methods. While the previous atom-wise approach uses sketch-map to visualize datasets and excluded hydrogen atoms, the proposed framework successfully separates compositions using information from all atoms together with relatively simple dimensionality reduction methods, such as MDS and Isomap.

\subsection{Benchmark in Data Selection}
\subsubsection{Analysis Procedure}
Beyond dataset visualization, similarity evaluation can play a key role in data selection for MLIP training. The objective is to construct a compact yet diverse training dataset from a large candidate pool. Since atomic feature vectors can be computed prior to DFT calculations, an effective similarity metric enables the diversity of candidate structures to be assessed before generating reference energies and forces. This makes similarity-based data selection a practical strategy for constructing diverse MLIP training datasets directly from unlabeled collections of candidate structures. We adopt similarity as the sole selection criterion to clearly demonstrate the intrinsic effect of our proposed methods, although combining it with other metrics, such as uncertainty estimates, is also feasible.

We employ farthest point sampling (FPS), a similarity-based data-selection strategy that utilizes our similarity evaluation method. The procedure for selecting the training dataset is as follows:

\begin{enumerate}
\item Initialize the training dataset with one randomly chosen data point from the candidate pool.
\item Select the data point from the candidate pool that maximizes its minimum distance or, equivalently, minimizes its maximum similarity to any data point in the current training dataset.
\item Repeat this process until the target dataset size is reached.
\end{enumerate}

We generate training datasets of increasing sizes using this FPS procedure with our proposed cluster-wise methods and two conventional data-wise and atom-wise methods. We also include a baseline in which the training dataset is selected randomly. We then train MLIPs on these datasets of varying sizes and evaluate their energy and force-prediction accuracy. Finally, we create learning curves with the number of training data points on the x-axis and the root mean square error (RMSE) on the y-axis to compare the data selection efficiency of each method. For the MLIP, we choose a Gaussian approximation potential (GAP) model \cite{bartok2010gaussian} implemented with Turbo-GAP code \cite{caro2019optimizing}.

\subsubsection{Target Datasets}
For this benchmark, we use two datasets. One is the amorphous carbon dataset \cite{deringer2017machine}, which consists solely of carbon and contains only amorphous structures. This composition eliminates chemical diversity, enabling a focused evaluation of the learning efficiency associated with the structural diversity.

\begin{figure}
\includegraphics[width=\textwidth]{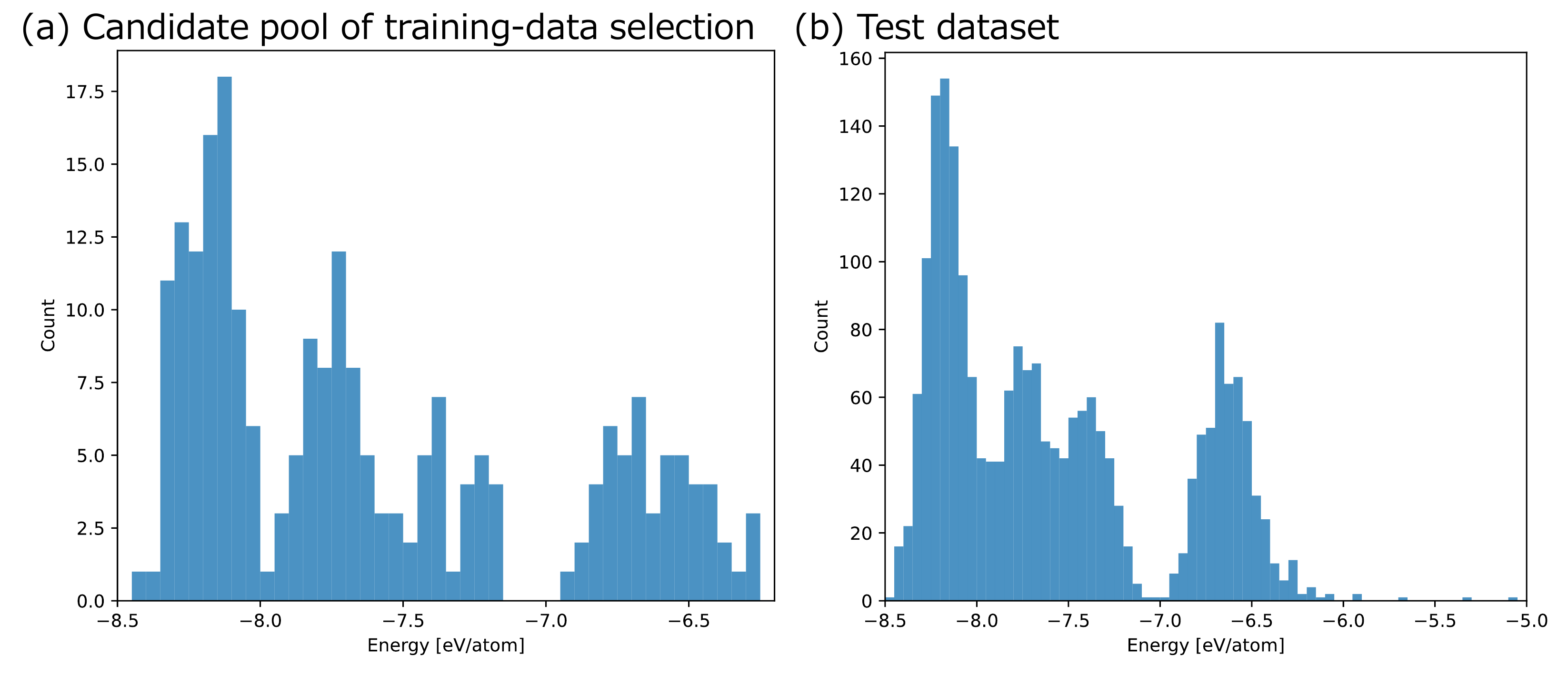}
\caption{Distribution of the potential energy per atom in the amorphous carbon dataset for (a) the candidate pool of training-data selection and (b) the test dataset.}
\label{fig:dataset_a_C}
\end{figure}

While the original dataset contains structures with varying numbers of atoms, we extract only structures containing 64 atoms to ensure consistent comparison conditions. From this subset, $2,168$ structures are used as the test dataset, while $225$ structures are used as the candidate pool for training-data selection. The potential-energy distributions of the candidate and test datasets, as shown in Fig.~\ref{fig:dataset_a_C} , are comparable, indicating that both datasets contain structures with similar levels of stability.

The performance of each data-selection strategy is evaluated using the same test dataset. The objective of this benchmark is to clarify how effectively similarity-based sampling can identify informative structures and improve learning efficiency.

The other dataset consists of germanium, antimony, and tellurium, hereafter referred to as the GST dataset \cite{GST_inpre}. The relative proportions of the three species vary among the structures, and each structure contains between $351$ and $363$ atoms, which are large numbers of atoms compared with those in the amorphous carbon dataset. Because the dataset exhibits both compositional and structural diversity, it provides a practical benchmark for evaluating the learning efficiency of the proposed framework.

\begin{figure}
\includegraphics[width=\textwidth]{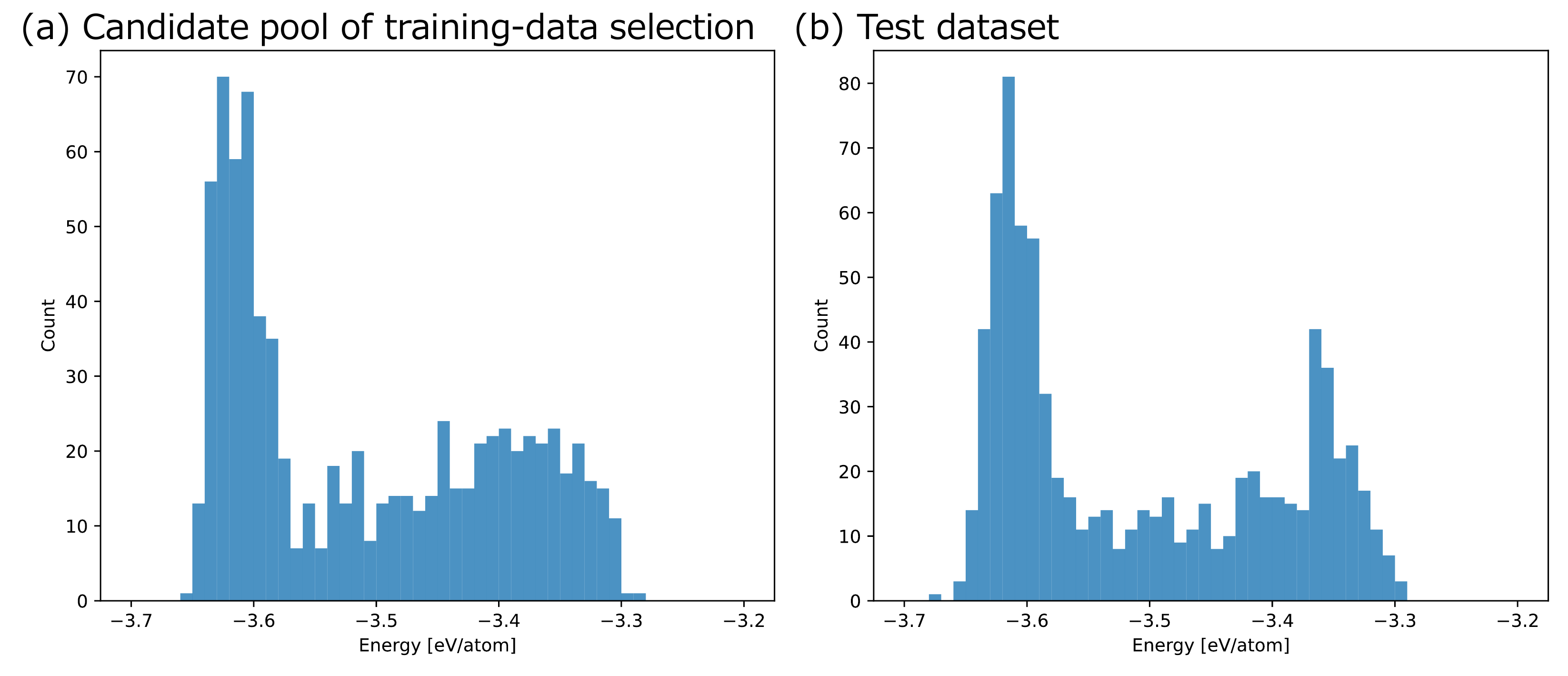}
\caption{Distribution of the potential energy per atom in the GST dataset for (a) the candidate pool of training-data selection and (b) the test dataset.}
\label{fig:dataset_gst}
\end{figure}

The dataset contains $1,600$ structures, of which $800$ are used as the candidate pool for training-data selection and the remaining $800$ as the test dataset. This separation enables us to evaluate the generalization performance of the resulting GAP models. As shown in Fig.~\ref{fig:dataset_gst}, the candidate and test datasets exhibit comparable potential-energy distributions, indicating that they contain structures with similar levels of stability.

\subsubsection{Benchmark on Carbon Dataset}
\begin{figure}
\includegraphics[width=\textwidth]{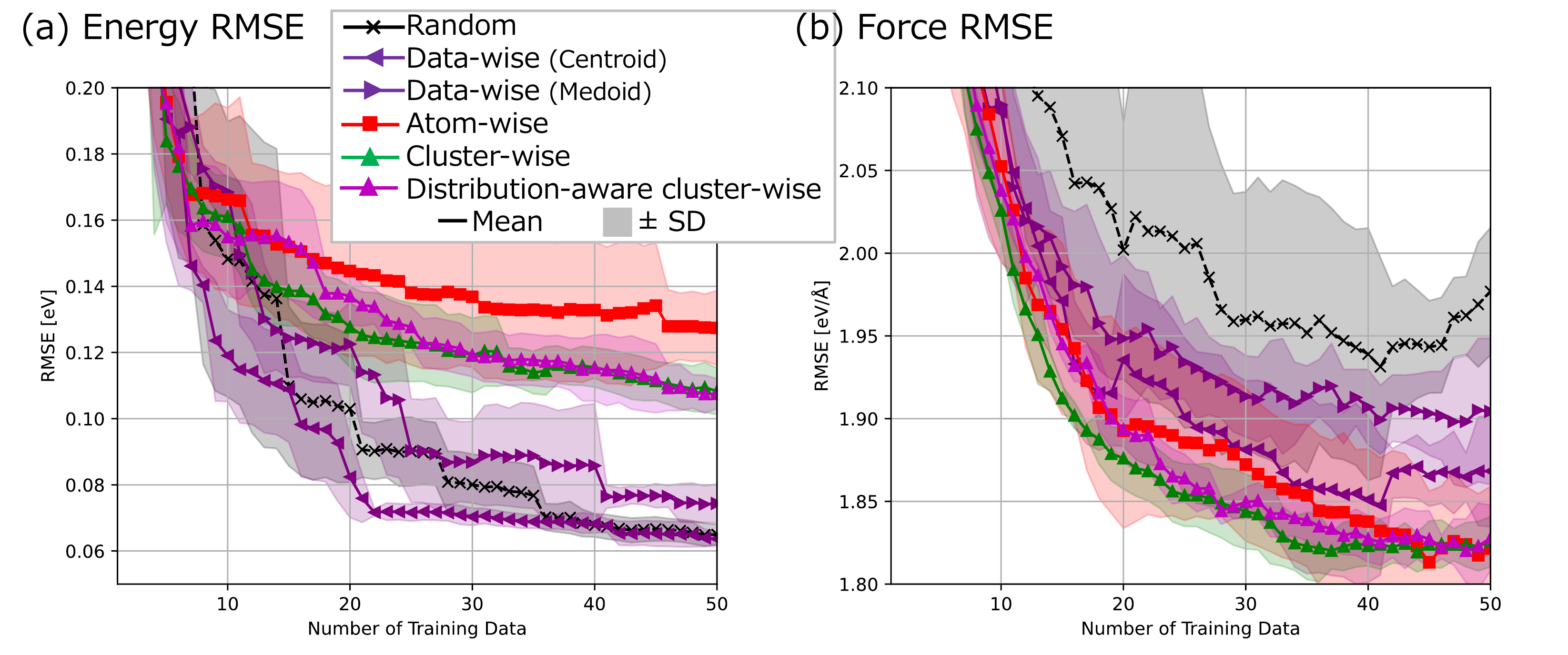}
\caption{Learning curves of (a) energy and (b) force RMSEs for the GAP models trained on amorphous carbon datasets constructed with different data-selection strategies. Both RMSEs are plotted as a function of the number of training data points. Each line represents the mean RMSE over $5$ independent trials, and the shaded areas denote the corresponding standard deviations.}
\label{fig:a_C_rmse}
\end{figure}

Fig.~\ref{fig:a_C_rmse} shows the learning curves of the GAP models trained on datasets constructed with each selection strategy. The lines with markers indicate the average RMSE over $5$ independent trials with different initial data points, while the shaded areas denote the corresponding standard deviations. The x-axis represents the number of training data points, and the y-axis represents the energy or force RMSE. Random sampling is shown by the black dashed line. The data-wise methods using centroid and medoid are shown in purple, the atom-wise method in red, the cluster-wise method in green, and the distribution-aware cluster-wise method in magenta. Detailed conditions of this benchmark are provided in Appendix~\ref{appendix:a-c}.

In terms of force RMSE, all similarity evaluation methods outperform random sampling. This indicates that similarity-based sampling is advantageous compared to random sampling for force prediction.

The cluster-wise method consistently achieves the lowest force RMSE across all training dataset sizes, demonstrating the highest learning efficiency. Furthermore, a notable difference is observed among the methods in terms of the stability of the learning process. The learning curve for the atom-wise method exhibits large variations across trials, resulting in a large standard deviation. In contrast, the cluster-wise methods show stable convergence with a remarkably small standard deviation.

The trend is different for the energy RMSE. Random sampling and the data-wise methods achieve lower energy RMSE values, whereas the cluster-wise, distribution-aware cluster-wise, and atom-wise methods consistently yield higher energy RMSE values. The origin of this discrepancy between energy and force prediction performance is discussed in the following section.

\begin{figure}
\includegraphics[width=\textwidth]{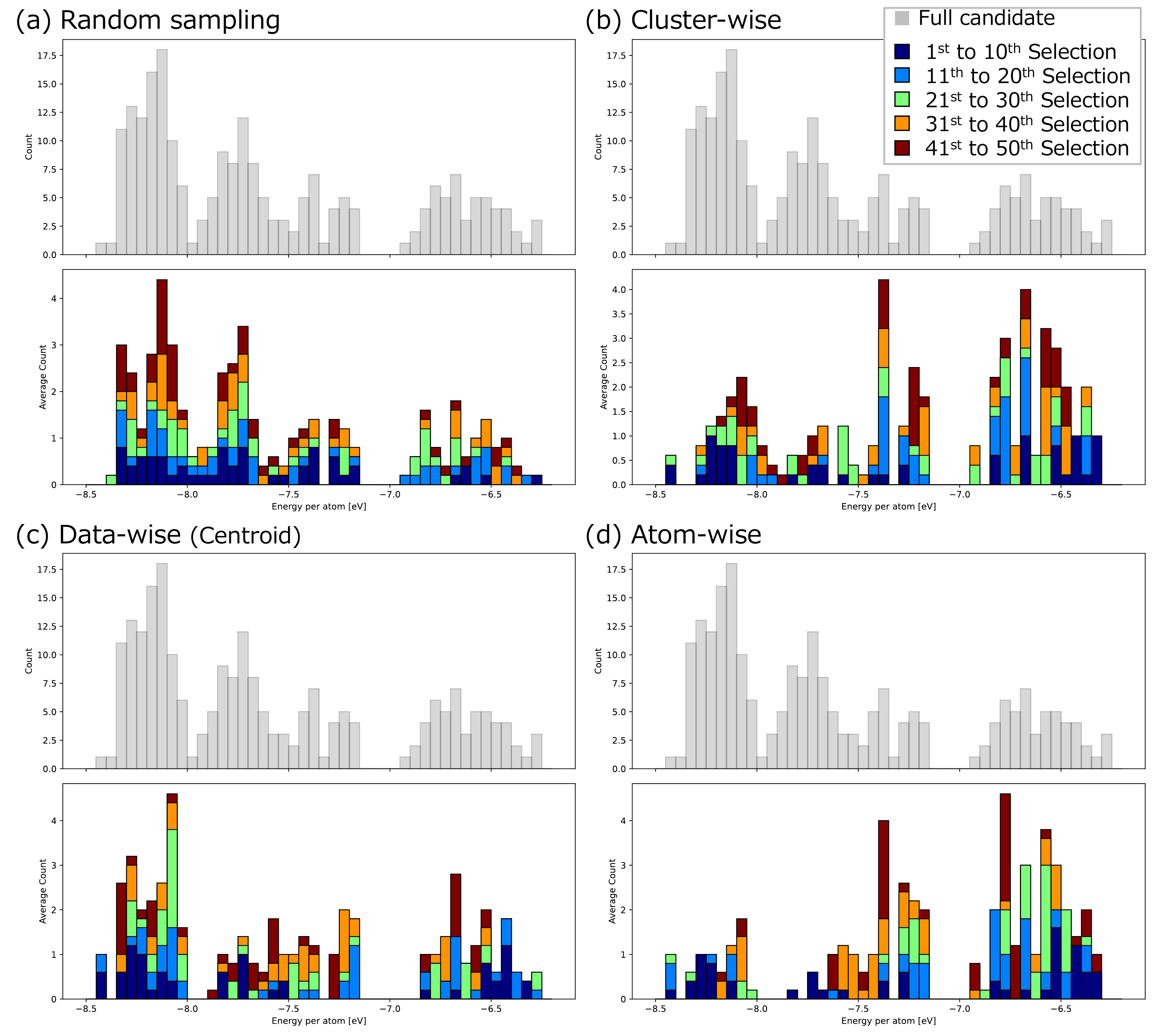}
\caption{Potential-energy distributions of the structures selected by (a) random, (b) cluster-wise, (c) data-wise, and (d) atom-wise sampling on the amorphous carbon dataset. The upper histograms show the distribution of the full candidate dataset, while the lower histograms show the ensemble-averaged distributions of the 50 selected structures obtained from $5$ independent trials.}
\label{fig:a_C_train}
\end{figure}

To understand the origin of the trends observed in Fig.~\ref{fig:a_C_rmse}, we examine the distributions of the data selected by each sampling strategy. Fig.~\ref{fig:a_C_train} shows the potential energy distributions of the selected training datasets when $50$ structures are selected. In each panel, the upper histogram represents the energy distribution of the full candidate dataset, while the lower histogram shows the energy distribution of the selected structures. The colors indicate the selection order, allowing the progression of the sampling process to be visualized.

As expected, the energy distribution obtained by random sampling closely follows that of the full candidate dataset. The data-wise method using the centroid representation produces a very similar distribution, even though the FPS algorithm is designed to maximize structural diversity. This observation suggests that the data-wise similarity measure provides limited discrimination between candidate structures, resulting in a sampling process that is effectively close to random sampling.

In contrast, both the atom-wise and cluster-wise methods preferentially select higher-energy structures. This behavior is consistent with the objective of similarity-based sampling, which seeks structurally diverse configurations that are often located farther away from the bottom of the potential energy surface. The atom-wise method exhibits the strongest bias toward high-energy structures, whereas the cluster-wise method maintains a more balanced coverage of both high- and low-energy regions.

\begin{figure}
\includegraphics[width=\textwidth]{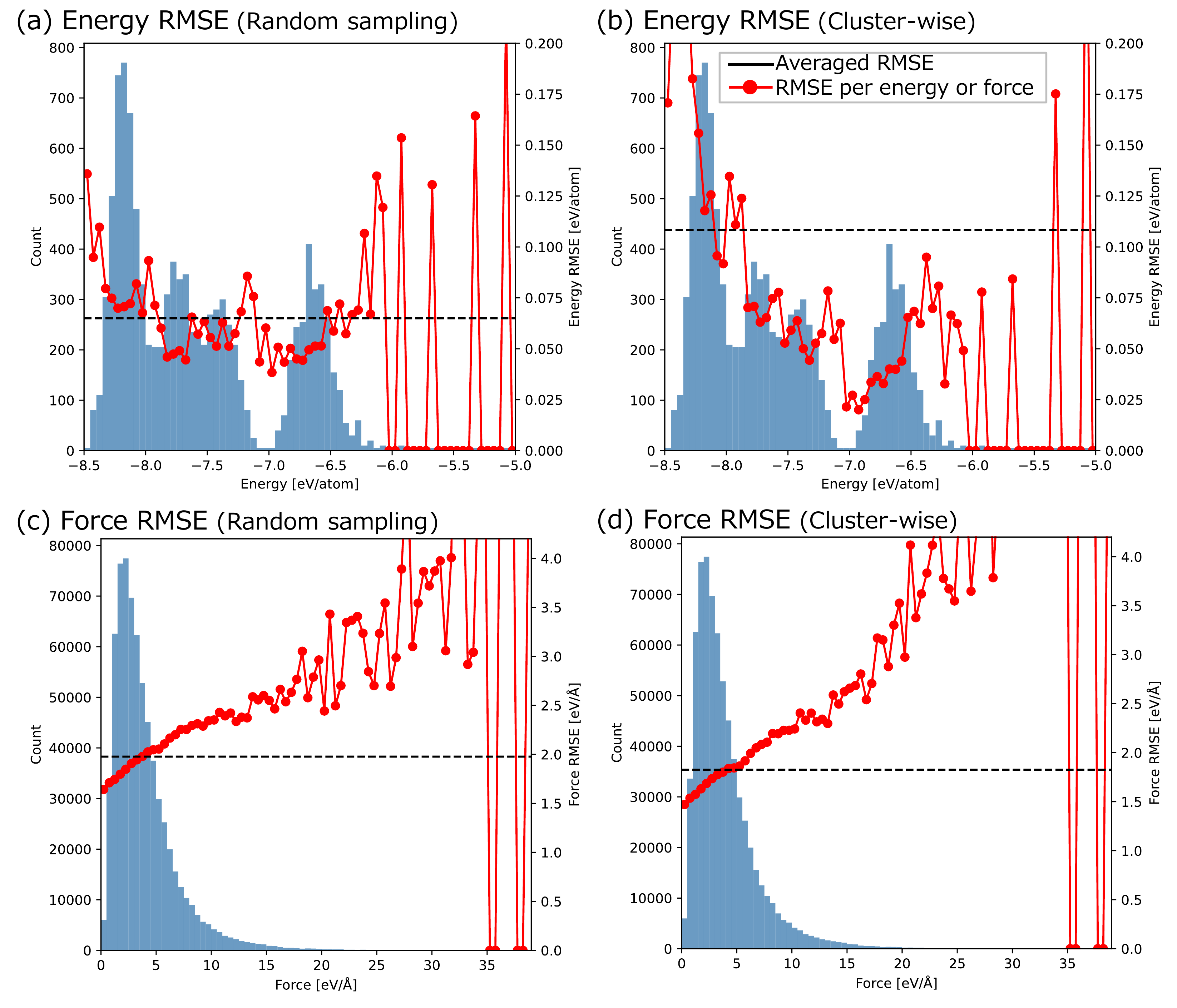}
\caption{Energy and force RMSEs as functions of the potential energy per atom and force magnitude for the amorphous carbon dataset. The histograms show the distribution of the test dataset, while the red lines indicate the corresponding RMSE values. (a) and (c) show the results obtained using random sampling, and (b) and (d) show the results obtained using cluster-wise sampling.}
\label{fig:a_C_test}
\end{figure}

Fig.~\ref{fig:a_C_test} (a, b) shows the energy RMSE of random sampling and the cluster-wise method. The cluster-wise method yields smaller prediction errors in the high-energy region, whereas random sampling performs better in the low-energy region where most structures are concentrated. Because the overall energy RMSE is dominated by these highly populated low-energy structures, the advantage of the cluster-wise method in the high-energy region is not reflected in the aggregate metric. However, this behavior may be beneficial for MD simulations. The cluster-wise method preferentially samples configurations that are located farther away from the bottom of the potential energy surface, thereby improving the coverage of diverse local environments. As a result, the trained GAP model is expected to remain reliable over a wider region of configuration space encountered during MD simulations.

Fig.~\ref{fig:a_C_test} (c, d) shows the force RMSE of random sampling and the cluster-wise method. The cluster-wise method consistently achieves lower force RMSE across the low-force region, which contains the majority of test data points. Consequently, the improvement in this densely populated region directly translates into a lower overall force RMSE.

\subsubsection{Benchmark on GST Dataset}
\begin{figure}
\includegraphics[width=\textwidth]{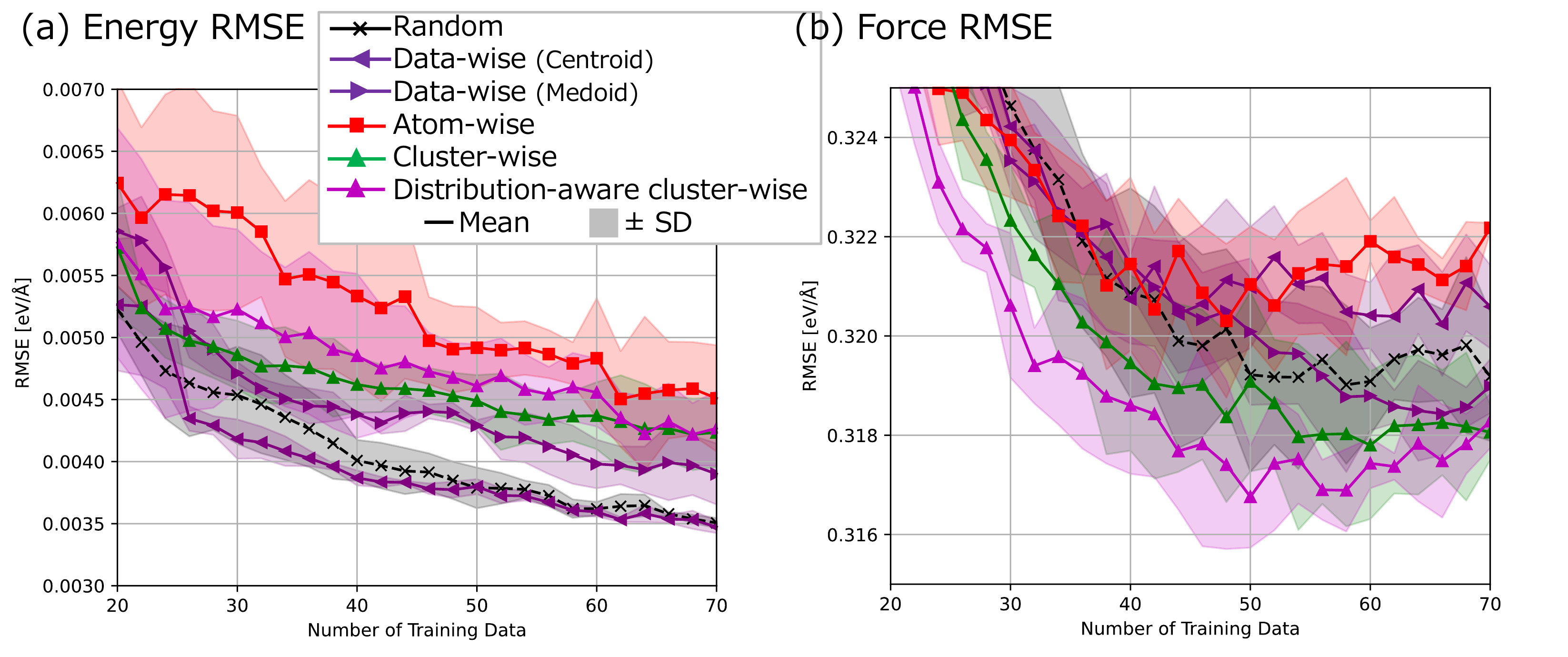}
\caption{Learning curves of (a) energy and (b) force RMSEs for the GAP models trained on GST datasets constructed with different data-selection strategies. Both RMSEs are plotted as a function of the number of training data points. Each line represents the mean RMSE over $3$ independent trials, and the shaded areas denote the corresponding standard deviations.}
\label{fig:gst_rmse}
\end{figure}

Fig.~\ref{fig:gst_rmse} shows the learning curves of the GAP models trained on datasets constructed with each selection strategy. The lines with markers indicate the average RMSE over $3$ independent trials with different initial data points, while the shaded areas denote the corresponding standard deviations. Detailed conditions of this benchmark are provided in Appendix~\ref{appendix:gst}.

In terms of force RMSE, the distribution-aware cluster-wise method achieves the lowest RMSE over much of the evaluated range of training dataset sizes. The standard cluster-wise method also exhibits favorable performance and generally outperforms random sampling. In contrast, the improvement obtained with the atom-wise method is limited as the training dataset size increases, and its force RMSE eventually becomes higher than that of random sampling. The performance of the data-wise method depends on the choice of representative vector: the centroid-based representation yields relatively low force RMSE values, whereas the medoid-based representation performs less favorably.

In terms of energy RMSE, random sampling and the data-wise methods using centroid achieve lower energy RMSE values, whereas the cluster-wise, distribution-aware cluster-wise, and atom-wise methods consistently yield higher energy RMSE values. This behavior is consistent with the trend observed for the amorphous carbon dataset.

\begin{figure}
\includegraphics[width=\textwidth]{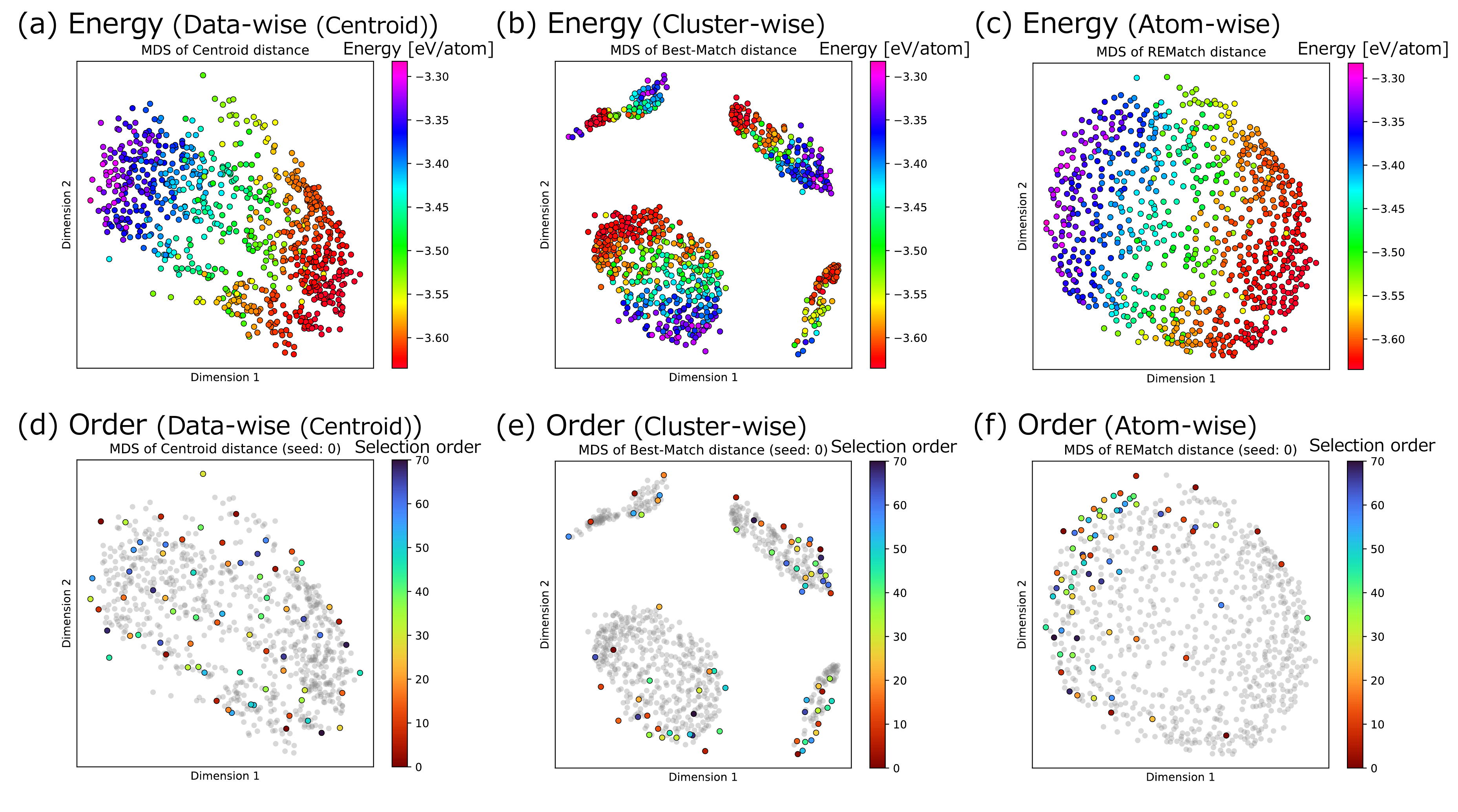}
\caption{Comparison of the MDS visualizations colored by (a, b, c) potential energy per atom and (d, e, f) selection order obtained using (a, d) the data-wise, (b, e) cluster-wise, and (c, f) atom-wise methods.}
\label{fig:gst_mds}
\end{figure}

To further investigate the sampling behavior underlying the learning-curve results, we visualize the GST dataset using MDS and examine the structures selected by each data-selection strategy. Fig.~\ref{fig:gst_mds} shows the MDS visualizations of the distance matrices derived from the proposed and baseline methods. The plots in the upper panels are colored according to their potential energy per atom, while the lower panels are colored according to their selection order. 

In all plots, the potential energy varies systematically across the MDS space, indicating that the corresponding distances derived from the data-wise, cluster-wise, and atom-wise methods capture structural differences associated with potential energy. However, the spatial distributions of the selected structures differ substantially among the methods.

For the data-wise method, the selected structures are broadly scattered throughout the embedding and include both low- and high-energy configurations. Although FPS is designed to select mutually dissimilar structures, this selection pattern is relatively close to random sampling. This result suggests that excessive compression into a single centroid limits the ability of the data-wise similarity measure to discriminate between structurally distinct configurations.

In contrast, the cluster-wise and atom-wise methods select data points located in the high-energy regions. This behavior indicates that these methods more effectively identify structurally distinct configurations, which tend to lie farther from the densely populated low-energy regions. The atom-wise method shows a particularly strong preference for high-energy structures, whereas the cluster-wise method also selects a substantial number of low-energy configurations, resulting in more balanced coverage of the energy range.

\begin{figure}
\includegraphics[width=\textwidth]{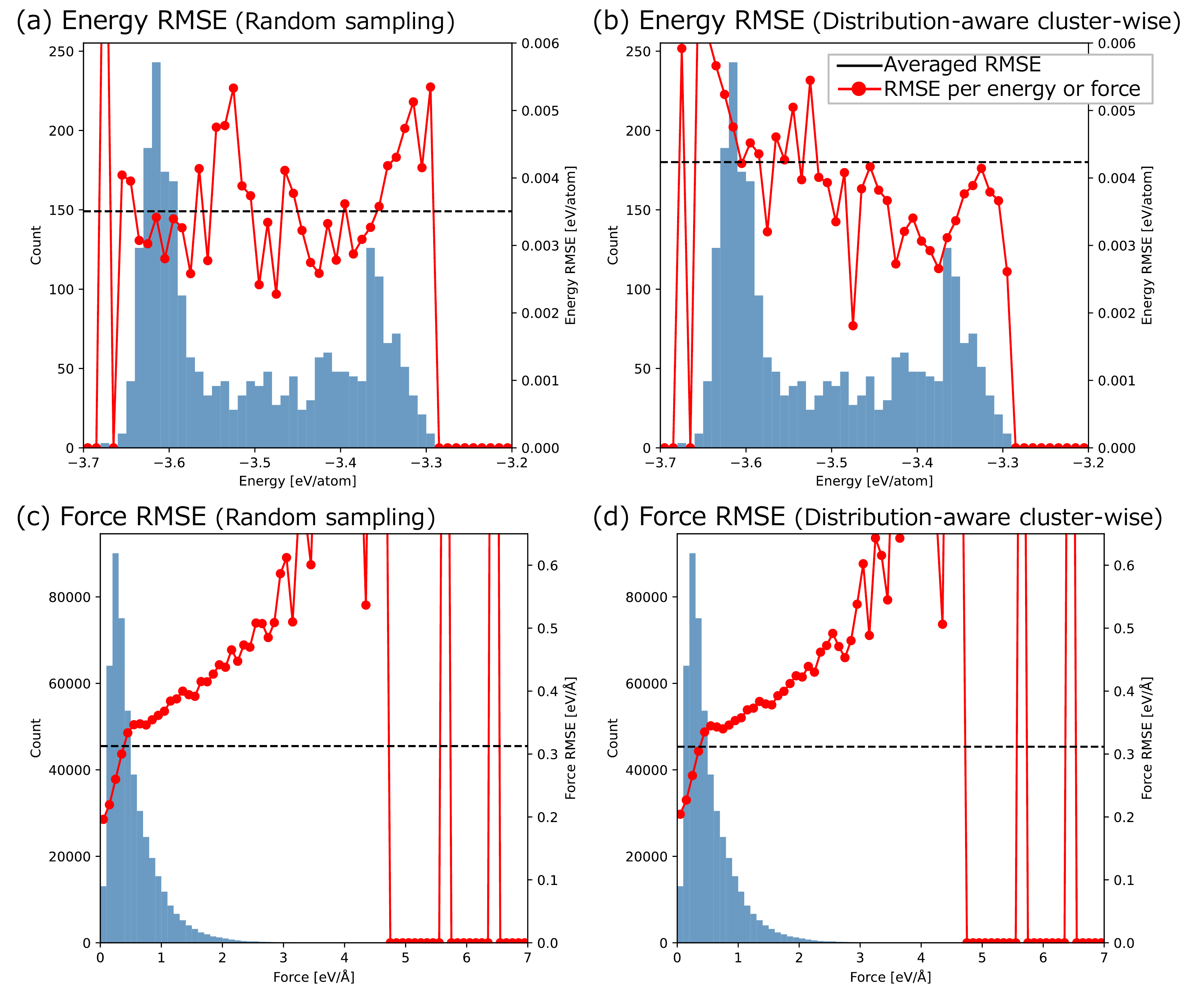}
\caption{Energy and force RMSEs as functions of the potential energy per atom and force magnitude for the GST dataset. The histograms show the distribution of the test dataset, while the red lines indicate the corresponding RMSE values. (a) and (c) show the results obtained using random sampling, and (b) and (d) show the results obtained using distribution-aware cluster-wise sampling.}
\label{fig:gst_test}
\end{figure}

Fig.~\ref{fig:gst_test} (a, b) shows the energy RMSE obtained using random sampling and the distribution-aware cluster-wise method. Similar to the amorphous carbon results, random sampling achieves lower errors in the low-energy region, whereas the distribution-aware cluster-wise method tends to perform better in parts of the higher-energy region. Because the aggregate energy RMSE is dominated by the low-energy region, this high-energy advantage is not reflected in the overall energy RMSE.

Fig.~\ref{fig:gst_test} (c, d) shows the force RMSE obtained using random sampling and the distribution-aware cluster-wise method. The distribution-aware cluster-wise method achieves slightly lower errors in the densely populated low-force region, leading to a lower overall force RMSE. The errors fluctuate considerably at larger force values because only a limited number of test data points are available in this region.

\subsubsection{Summary}
The benchmark results demonstrate that data selection based on the proposed cluster-wise methods is effective for improving force-prediction accuracy and reducing variability across independent trials, compared with the baseline sampling strategies. This advantage was observed for both the amorphous carbon dataset, which contains only a single chemical species, and the GST dataset, which contains germanium, antimony, and tellurium in varying proportions. These results indicate that the proposed framework is applicable to datasets exhibiting structural diversity alone as well as those exhibiting both structural and compositional diversity.

The different similarity evaluation methods produce distinct sampling behaviors. The data-wise method produced a selection pattern similar to random sampling, suggesting that compression into a single representative vector provides insufficient resolution for the data selection task. In contrast, the atom-wise method strongly favored structurally distinct, high-energy configurations. The cluster-wise methods also preferentially selected high-energy configurations but retained a larger proportion of low-energy structures, thereby providing a more balanced coverage of the potential-energy range. This balance likely contributes to their favorable force-prediction performance.

Notably, these sampling patterns are obtained without using potential energy or any other target property as a selection criterion. Instead, the proposed methods identify diverse configurations solely from differences in their local atomic environments. Because these atomic descriptors can be computed from the atomic structures before reference energies and forces are generated by DFT, the cluster-wise framework can be applied to unlabeled candidate structures. It therefore provides a practical approach for selecting diverse structures prior to costly reference calculations and for constructing efficient training datasets for MLIPs.

\section{Discussion}
In this study, we developed a similarity evaluation method based on clustering to overcome the trade-off between representational fidelity and computational cost in assessing the similarity between data points in materials datasets. This method first uses $k$-medoids clustering to summarize the diverse atomic environments within each data point into a small and fixed number of representative vectors. Subsequently, the similarity score between data pairs is computed by solving an optimal matching problem between the corresponding sets of representative vectors.

We validated the ability of similarity evaluation methods to distinguish chemically different data points through qualitative evaluations on the QM7b and SHIFTML-molfrags datasets. Visualization of the distance matrix using MDS and Isomap revealed that both cluster-wise methods achieved clearer compositional separation according to the chemical composition than the data-wise and atom-wise methods. These results indicate that the proposed standard and distribution-aware cluster-wise methods can sensitively capture chemically meaningful differences. In addition, the standard cluster-wise method also achieved an approximately $50$-fold speedup in this validation on the QM7b dataset. 

Subsequently, we evaluated the practical utility of similarity evaluation methods through data selection benchmarks on the amorphous carbon and GST datasets. The results demonstrated that the standard and distribution-aware cluster-wise methods provide favorable force-prediction accuracy and reduced variability across independent GAP training runs compared with the baseline methods. These results highlight the effectiveness of the proposed cluster-wise method for similarity-based selection of diverse MLIP training data.

\section{Methods}
This section details and contrasts four distinct similarity evaluation methods, including our proposed approaches, for quantifying differences in atomic environments between two data points. 

\subsection{Data-wise Similarity Evaluation}
This method summarizes the set of feature vectors for each data point into a single representative vector. A similarity score is then calculated from the two representative vectors. 

To represent the local atomic environments of each atom, a descriptor is used to transform the positions and chemical species of its neighboring atoms into a fixed-size numerical feature vector. While numerous descriptors have been developed, this study employs the Smooth Overlap of Atomic Positions (SOAP) descriptor \cite{bartok2010gaussian,bartok2013representing}. A key advantage of SOAP is its invariance under translation, rotation, and permutation of chemically equivalent atoms. Specifically, we use the SOAP-Turbo variant, selected for its computational efficiency and compact representation \cite{caro2019optimizing}.

The sets of normalized SOAP feature vectors for data points A and B are denoted by $\mathbf{P}$ and $\mathbf{Q}$, respectively, and defined as follows: 
\begin{equation}
\label{SOAP_data_A}
\mathbf{P}=\left[\begin{array}{cccc}
\mathbf{p}_{\mathrm{1}} & \mathbf{p}_{\mathrm{2}} & \cdots & \mathbf{p}_{n_{\mathrm{A}}}
\end{array}\right]^{\mathrm{T}},
\end{equation}
\begin{equation}
\label{SOAP_data_B}
\mathbf{Q}=\left[\begin{array}{cccc}
\mathbf{q}_{\mathrm{1}} & \mathbf{q}_{\mathrm{2}} & \cdots & \mathbf{q}_{n_{\mathrm{B}}}
\end{array}\right]^{\mathrm{T}},
\end{equation}
where $\mathbf{p}_{i}$ and $\mathbf{q}_{j}$ are the SOAP feature vectors for the $i$~th atom in data point A and the $j$~th atom in data point B, respectively, and $n_{\mathrm{A}}$ and $n_{\mathrm{B}}$ are the numbers of atoms in data points A and B.

A representative vector for each data point can be defined in several ways, such as the mean vector or a medoid. When the mean vectors are used as the representative vectors, those of data points A and B are denoted by $\mathbf{\bar{p}}$ and $\mathbf{\bar{q}}$. The similarity score between two data points is then calculated from the inner product of their representative vectors as follows:
\begin{equation}
\label{data-wise:k}
k_{\mathrm{A},\mathrm{B}}=\left(\mathbf{\bar{p}}\cdot\mathbf{\bar{q}}\right)^{\zeta},
\end{equation}
where ${\zeta}$ is a hyperparameter that controls the sharpness of the similarity score. In this study, ${\zeta}$ is set to $2$.

The data-wise method is computationally inexpensive as it only requires the calculation of one representative vector for each data point and one inner product for each data pair. However, it inherently compresses the information of all atomic environments into a single representative vector. Particularly for data points containing diverse atomic environments or a large number of atoms, this compression can lead to an inadequate representation of the data point and result in a low-fidelity similarity score.

\subsection{Atom-wise Similarity Evaluation}
This method, proposed by De et al. \cite{de2016comparing}, avoids any form of data compression. Instead, it computes a data-level similarity score from a comprehensive comparison of all SOAP feature vectors from the two data points.

Since this method operates directly on the individual atomic feature vectors, the calculation of the representative vector is not required. Consequently, the similarity score between data points is formed using the entire set of feature vectors as follows:
\begin{equation}
\label{atom_wise:k}
k_{\mathrm{A},\mathrm{B}}=\sum_{i=1,j=1}^{n_{\mathrm{A}},n_{\mathrm{B}}} X_{i,j}\left(\mathbf{p}_{i}\cdot\mathbf{q}_{j}\right)^{\zeta},
\end{equation}
where $X_{i,j}$ represents the contribution of the pairing between the feature vector of the $i$-th atom in data point A and that of the $j$-th atom in data point B. The similarity score is normalized to the range from 0 to 1 by constraining the matching matrix $\mathbf{X}$, composed of these $X_{i,j}$ elements, by a doubly stochastic matrix as follows:
\begin{equation}
\label{atom_wise:const1}
\mathrm{s.t.} \sum_{j=1}^{n_{\mathrm{B}}} X_{i,j} = \frac{1}{n_{\mathrm{A}}} \quad \text{for }i=1,\dots,n_{\mathrm{A}},
\end{equation}
\begin{equation}
\label{atom_wise:const2}
\sum_{i=1}^{n_{\mathrm{A}}} X_{i,j} = \frac{1}{n_{\mathrm{B}}} \quad \text{for }j=1,\dots,n_{\mathrm{B}}.
\end{equation}
Because the SOAP descriptor treats different chemical species as independent channels, the similarity between atomic environments associated with different central chemical species is set to zero in this study. Consequently, pairings between different chemical species do not contribute to the similarity score.

There are two promising strategies for obtaining $\mathbf{X}$ that satisfy these constraints. One is an exclusive pairing strategy, which mandates a strict one-to-one correspondence between the atoms from each data point. This approach is applicable only when the number of atoms in both data points is identical ($n_{\mathrm{A}}=n_{\mathrm{B}}$), and therefore, it is rarely applicable to the atom-wise method. The optimal $\mathbf{X}$ is calculated as follows:
\begin{equation}
\label{atom_wise:best-match}
\mathbf{\hat{X}}=\underset{\mathbf{X}}{\arg\max}\sum_{i=1,j=1}^{n_\mathrm{A}} X_{i,j}\left(\mathbf{p}_{i}\cdot\mathbf{q}_{j}\right)^{\zeta},
\end{equation}
where $\mathbf{X}$ is a scaled permutation matrix that has exactly one entry of $1/n_{\mathrm{A}}$ in each row and each column and zeros elsewhere. This optimization problem can be solved using the Hungarian algorithm \cite{kuhn1955hungarian}. Hereafter, this exclusive pairing strategy is referred to as the best-match method.

The other is a non-exclusive pairing strategy, which allows atoms to be reused in multiple combinations. This is formulated as an entropy-regularized optimal transport problem:
\begin{equation}
\label{atom_wise:REMatch}
\mathbf{\hat{X}}=\underset{\mathbf{X}}{\arg\min}\sum_{i=1,j=1}^{n_{\mathrm{A}},n_{\mathrm{B}}} X_{i,j}\left(1-\left(\mathbf{p}_{i}\cdot\mathbf{q}_{j}\right)^{\zeta}\right)+\gamma \sum_{i=1,j=1}^{n_{\mathrm{A}},n_{\mathrm{B}}} X_{i,j}\log X_{i,j},
\end{equation}
where $\mathbf{X}$ satisfies the constraints in Eqs.~\eqref{atom_wise:const1} and \eqref{atom_wise:const2}. In contrast to the discrete assignments of the best-match method, $X_{i,j}$ can take on continuous fractional values. Here, $\gamma$ is a regularization parameter controlling the smoothness of the solution. A sufficiently small $\gamma$ yields results practically identical to the best-match method. Therefore, we use a small value of $\gamma$ as the default setting. Conversely, a large $\gamma$ can cause the regularization term to become dominant, reducing the effect of the similarity represented by the inner product.

This optimization problem can be solved efficiently using the Sinkhorn algorithm \cite{sinkhorn1967concerning,cuturi2013sinkhorn}. Hereafter, this non-exclusive pairing strategy is referred to as the regularized-entropy match (REMatch) method. The REMatch method is the default choice for atom-wise similarity evaluation because it can compare data points containing different numbers of atoms, whereas the best-match method requires equal atom counts.

\subsection{Cluster-wise Similarity Evaluation}
This method computes a similarity score through a multi-step process. First, $k$-medoids clustering is applied to the set of all feature vectors of each data point, partitioning them into a predefined number of clusters $c$. The medoid of each cluster is then selected as its representative vector. This step effectively reduces the large, variable-size set of feature vectors into a small, fixed-size set of representative vectors. For two data points A and B, the resulting sets of representative vectors are denoted by $\mathbf{P}^{*}$ and $\mathbf{Q}^{*}$, respectively:
\begin{equation}
\label{SOAP_data_A_clu}
\mathbf{P}^{*}=\left[\begin{array}{cccc}
\mathbf{p}_{\mathrm{1}}^{*} & \mathbf{p}_{\mathrm{2}}^{*} & \cdots & \mathbf{p}^{*}_{c}
\end{array}\right]^{\mathrm{T}},
\end{equation}
\begin{equation}
\label{SOAP_data_B_clu}
\mathbf{Q}^{*}=\left[\begin{array}{cccc}
\mathbf{q}_{\mathrm{1}}^{*} & \mathbf{q}_{\mathrm{2}}^{*} & \cdots & \mathbf{q}^{*}_{c}
\end{array}\right]^{\mathrm{T}}.
\end{equation}
Second, the similarity score between data points is calculated by applying the best-match method to these sets of representative vectors:
\begin{equation}
\label{cluster_wise:k}
k_{\mathrm{A},\mathrm{B}}=\sum_{i=1,j=1}^{c} X_{i,j}\left(\mathbf{p}_{i}^{*}\cdot\mathbf{q}_{j}^{*}\right)^{\zeta},
\end{equation}
where $\mathbf{X}$ is computed by applying the previously described best-match method shown in Eq.~\eqref{atom_wise:best-match} to $\mathbf{P}^{*}$ and $\mathbf{Q}^{*}$. Accordingly, $\mathbf{X}$ is a scaled permutation matrix, and $k_{\mathrm{A},\mathrm{B}}$ represents the average similarity between the matched representative vectors.

A key advantage of this cluster-wise method is that it enables the computationally efficient best-match method to be applied to data points containing different numbers of atoms, provided that the same number of clusters is used for both. This makes the best-match method an effective default choice for this approach.

The data-wise, atom-wise, and cluster-wise methods represent different trade-offs between computational cost and information loss. The data-wise and atom-wise methods define the two extremes of this spectrum: the former prioritizes maximum efficiency by aggregating all information into a single representative vector, while the latter aims for maximum fidelity by retaining all feature vectors. Our cluster-wise approach, including the distribution-aware cluster-wise method introduced in the following section, is designed as a balanced solution that achieves both high fidelity and computational efficiency. 

\subsection{Distribution-aware Cluster-wise Similarity Evaluation}
We further propose a distribution-aware cluster-wise similarity evaluation method that incorporates statistical information about the feature vectors within each cluster. The procedure is identical to that of the standard cluster-wise method up to the determination of the matching matrix. For each cluster, we additionally compute the covariance matrix of the feature vectors assigned to that cluster.

Here, we approximate each cluster by a multivariate Gaussian distribution whose mean is given by the cluster medoid and whose covariance matrix is calculated from the deviations of the feature vectors relative to that medoid:
\begin{equation}
\label{SOAP_data_A_clu_ws}
\mathcal{G}_{i}^{\mathrm{A}}=\mathcal{N}\left(\mathbf{p}_{i}^{*},\mathbf{\Sigma}_{i}^{\mathrm{A}}\right),
\end{equation}
\begin{equation}
\label{SOAP_data_B_clu_ws}
\mathcal{G}_{j}^{\mathrm{B}}=\mathcal{N}\left(\mathbf{q}_{j}^{*},\mathbf{\Sigma}_{j}^{\mathrm{B}}\right),
\end{equation}
where $\mathbf{\Sigma}_{i}^{\mathrm{A}}$ and $\mathbf{\Sigma}_{j}^{\mathrm{B}}$ denote the covariance matrices of cluster $i$ in data point A and cluster $j$ in data point B. The overall distance between a data pair is then defined as the weighted average of the 2-Wasserstein distances $d_{\mathrm{A},\mathrm{B}}^{\mathrm{WS}}$ between the matched cluster distributions:
\begin{equation}
\label{cluster_wise_ws:d_origin}
d_{\mathrm{A},\mathrm{B}}^{\mathrm{WS}}=\sum_{i=1,j=1}^{c}X_{i,j}\left(\|\mathbf{p}_{i}^{*}-\mathbf{q}_{j}^{*}\|_{2}^{2}+\mathrm{Tr}\left(\mathbf{\Sigma}_{i}^{\mathrm{A}}+\mathbf{\Sigma}_{j}^{\mathrm{B}}-2\left({\mathbf{\Sigma}_{i}^{\mathrm{A}}}^{\frac{1}{2}}{\mathbf{\Sigma}_{j}^{\mathrm{B}}}{\mathbf{\Sigma}_{i}^{\mathrm{A}}}^{\frac{1}{2}}\right)^{\frac{1}{2}}\right)\right)^{\frac{1}{2}}.
\end{equation}
Since the above equation involves matrix multiplications, calculating $d_{\mathrm{A},\mathrm{B}}^{\mathrm{WS}}$ for all data pairs is not realistic. Therefore, to improve computational efficiency, we approximate each covariance matrix as a diagonal matrix:
\begin{equation}
\label{SOAP_data_A_clu_ws_cov}
\mathbf{\Sigma}_{i}^{\mathrm{A}}=\mathrm{diag}\left[\left(\mathbf{\sigma}_{i}^{\mathrm{A}}\right)^{2}\right],
\end{equation}
\begin{equation}
\label{SOAP_data_B_clu_ws_cov}
\mathbf{\Sigma}_{j}^{\mathrm{B}}=\mathrm{diag}\left[\left(\mathbf{\sigma}_{j}^{\mathrm{B}}\right)^{2}\right],
\end{equation}
where $\mathbf{\sigma}_{i}^{\mathrm{A}}$ and $\mathbf{\sigma}_{j}^{\mathrm{B}}$ denote the standard deviation vectors of data points A and B, respectively. Under this assumption, the 2-Wasserstein distance can be simplified as follows:
\begin{equation}
\label{cluster_wise_ws:d_final}
d_{\mathrm{A},\mathrm{B}}^{\mathrm{WS}}=\sum_{i=1,j=1}^{c}X_{i,j}\left(\|\mathbf{p}_{i}^{*}-\mathbf{q}_{j}^{*}\|_{2}^{2}+\|\mathbf{\sigma}_{i}^{\mathrm{A}}-\mathbf{\sigma}_{j}^{\mathrm{B}}\|_{2}^{2}\right)^{\frac{1}{2}}.
\end{equation}
This equation incorporates not only the differences between cluster representative vectors but also the differences in the distributions of feature vectors within each cluster, while maintaining computational efficiency. Because $\mathbf{X}$ is obtained by the best-match method, the resulting data-level distance can be interpreted as an average of the 2-Wasserstein distances between matched clusters.

\backmatter


\bmhead{Data Availability}
The QM7b, PET-MAD, and amorphous carbon datasets analyzed in this study are available from the sources cited in the manuscript. The GST dataset will also be made publicly available.

\bmhead{Code Availability}
The custom code used to implement the cluster-wise similarity measures and reproduce the analyses will be made publicly available in a version-controlled repository.

\bmhead{Acknowledgments}
Y. I. acknowledges Prof. Miguel Caro's group for their hospitality during his research visit to Aalto University, Finland.

\bmhead{Author Contributions}
Y.I. conceived the study, developed and implemented the similarity evaluation methods, performed the calculations, analyzed the results, and wrote the original manuscript. M.A.C. supervised the study, contributed to the interpretation of the results, provided computational resources, and reviewed and edited the manuscript.

\bmhead{Competing Interests}
Y.I. is an employee of Fujitsu Limited. M.A.C. declares no financial or non-financial competing interests.

\begin{appendices}

\section{Appendix}\label{appendix}
\subsection{Technical Setting}
This appendix summarizes the descriptor parameters and computational settings used for the qualitative validation and data-selection benchmarks. 

\subsubsection{Validation on QM7b Dataset}\label{appendix:qm7b}
For the QM7b dataset, the SOAP descriptors were computed using a cutoff radius of $4~\AA$, with $8$ radial basis functions and $8$ angular basis functions. For the atom-wise method, the regularization term was set to $0.1$. The maximum number of iterations was set to $10,000$, and the convergence criterion was defined as a residual below $1.0 \times 10^{-4}$.

\subsubsection{Validation on SHIFTML-molfrags Dataset}\label{appendix:shiftml}
For the SHIFTML-molfrags dataset, the SOAP descriptors were computed using a cutoff radius of $5~\AA$, with $10$ radial basis functions and $8$ angular basis functions. For the atom-wise method, the regularization term was set to $0.01$. The maximum number of iterations was set to $10,000$, and the convergence criterion was defined as a residual below $1.0 \times 10^{-4}$.

\subsubsection{Benchmark on Carbon Dataset}\label{appendix:a-c}
For the amorphous carbon dataset, the SOAP descriptors were computed using a cutoff radius of $5~\AA$, with $10$ radial basis functions and $8$ angular basis functions. For the atom-wise method, the regularization term was set to $0.01$. The maximum number of iterations was set to $10,000$, and the convergence criterion was defined as a residual below $1.0 \times 10^{-4}$.

The GAP model consisted of two-body and SOAP-Turbo descriptor terms. The cutoff radius and the numbers of radial and angular basis functions were set to the same values as those used for similarity evaluation. The numbers of sparse points were set to $20$ and $1,000$ for the two-body and SOAP-Turbo terms, respectively.

\subsubsection{Benchmark on GST Dataset}\label{appendix:gst}
For the GST dataset, the SOAP descriptors were computed using a cutoff radius of $5.5~\AA$, with $8$ radial basis functions and $8$ angular basis functions. For the atom-wise method, the regularization term was set to $0.01$. The maximum number of iterations was set to $10,000$, and the convergence criterion was defined as a residual below $1.0 \times 10^{-4}$.

The GAP model consisted of two-body and SOAP-Turbo descriptor terms. The cutoff radius and the numbers of radial and angular basis functions were set to the same values as those used for similarity evaluation. The numbers of sparse points were set to $20$ and $1,500$ for the two-body and SOAP-Turbo terms, respectively.




\end{appendices}


\bibliography{references}

\end{document}